\documentclass{emulateapj}
\usepackage{apjfonts}

\newcommand{\etal}{et al.~}
\def\gsim{\lower 2pt \hbox{$\, \buildrel {\scriptstyle >}\over
{\scriptstyle \sim}\,$}}
\def\lsim{\lower 2pt \hbox{$\, \buildrel {\scriptstyle <}\over
{\scriptstyle \sim}\,$}}

\def\chandra{{\sl Chandra~}}
\def\xmm{{\sl XMM-Newton}}
\def\cvi{C~{\scriptsize VI}}

\def\ciii{C~{\scriptsize III}}

\def\nvii{N~{\scriptsize VII}}

\def\oviii{O~{\scriptsize VIII}}
\def\ovii{O~{\scriptsize VII}}
\def\ovi{O~{\scriptsize VI}}

\def\neix{Ne~{\scriptsize IX}}

\shortauthors{Yao \etal}
\shorttitle{The Dearth of the Warm-Hot CGM}

\slugcomment{\em Accepted for publication in the  Astrophysical Journal}

\begin{document}

\title{The Dearth of Chemically Enriched Warm-Hot Circumgalactic Gas}
\author{Y. Yao\altaffilmark{1},
	Q. D. Wang\altaffilmark{2}, 
	S. V. Penton\altaffilmark{1},
	T. M. Tripp\altaffilmark{2}, 
	J. M. Shull\altaffilmark{1}, and 
	J. T. Stocke\altaffilmark{1}
}
\altaffiltext{1}{Center for Astrophysics and Space Astronomy,
Department of Astrophysical and Planetary Sciences,
University of Colorado, 389 UCB, Boulder, CO 80309; 
Yangsen.Yao, Steven.Penton, 
Michael.Shull, John.Stocke@colorado.edu}
\altaffiltext{2}{Department of Astronomy, University of Massachusetts, 
  Amherst, MA 01003; wqd, tripp@astro.umass.edu}

\begin{abstract}

The circumgalactic medium (CGM) around galaxies is believed to record 
various forms of galaxy feedback and contain a significant portion of
the ``missing baryons'' of individual dark matter halos. However, clear 
observational evidence for the existence of the hot CGM is still absent. We
use intervening galaxies along 12 background AGNs as tracers
to search for X-ray 
absorption lines produced in the corresponding CGM. Stacking \chandra grating 
observations with respect to galaxy groups and different luminosities 
of these intervening galaxies, we obtain spectra with 
signal-to-noise ratios of 46-72 per 20-m\AA\ spectral bin 
at the expected \ovii\ K$\alpha$ line. We find no detectable 
absorption lines of \cvi, \nvii, \ovii, \oviii, or \neix. The high 
spectral quality allows us to tightly constrain upper limits to the 
corresponding ionic column densities (in particular
$\log[N{\rm_{OVII}(cm^{-2})}]\le$14.2--14.8). 
These nondetections are inconsistent with the Local Group hypothesis of
the X-ray absorption lines at $z\simeq0$ commonly observed in the spectra of 
AGNs. These results indicate that the putative CGM in the temperature range 
of $10^{5.5}-10^{6.3}$ K may not be able to account for the missing baryons
unless the metallicity is less than 10\% solar.

\end{abstract}


\section{Introduction }
\label{sec:intro}

Modern simulations are converging on the formation and evolution of
the dark galactic halos as well as the large-scale intergalactic 
structure of the Universe (e.g., \citealt{nav96, she01, cen99, dav99, spr05}).
However, serious difficulties are present in 
reproducing the visible
parts of galaxies, including the mass/luminosity
functions at low and high masses, the angular momentum distribution
of galactic disks, and the supermassive black hole (SMBH) and bulge-mass
relationship (e.g., \citealt{gil08, pri09, cat09}).
For instance, the ratio of the visible baryonic mass to the gravitational
mass of all galaxies is 3-10 times less than the cosmic value 
($0.167\pm0.006$) inferred from the {\sl Wilkinson Microwave Anisotropy Probe} 
(\citealt{kom09}; also see \citealt{bre09} and references therein);
the discrepancy seems to be more severe in less massive galaxies
than in more massive ones (e.g., \citealt{hoe05, mcg08}).
This missing link between the intergalactic medium (IGM), dark
halos, and visible galaxies is attributed largely to our poor 
understanding of the complex ``gastrophysics'', in particular the 
coupling between gas and galaxy feedback (e.g., \citealt{dav07}).

Galaxy feedback comes in many forms. Starbursts are known to generate 
galactic winds, which
must be chiefly responsible for the chemical enrichment and 
non-gravitational heating of the IGM at high redshifts 
(e.g., \citealt{mac99}). 
Nuclear starbursts are also believed to be intimately related to the 
formation and growth of the SMBHs,
which are an additional source of feedback, especially in massive 
galaxies (e.g., \citealt{str02, swa06}).
While the radiation pressure in a high-accretion phase can be a key 
driving force of the winds, the mechanical energy injection in
a radiatively inefficient accretion phase may balance the 
cooling of the surrounding gas. Feedback from stars, even from evolved 
ones alone (e.g., Type Ia supernovae), can play an essential role in 
maintaining hot gaseous halos around galaxies \citep{tang09a, tang09b}.
Cosmic-ray pressure can also play an important role in 
driving large-scale galactic outflows \citep{eve08}.
The relative importance of these various forms of the galaxy feedback, 
however, remains very unclear. Direct observational constraints on the 
physical process and history of the galaxy feedback are thus badly needed.

A potentially effective approach is to observe the circumgalactic 
medium (CGM) around nearby galaxies.
On scales from a few tens of kpc up to $\sim 1$ Mpc, the CGM includes
the gas that has been significantly affected by the galaxy 
feedback but is outside of the boundaries of the traditionally-known 
stellar and multi-phase interstellar components of such galaxies. 
It is quite possible that the CGM around the galaxies contains 
their ``missing'' baryons (e.g., \citealt{deh98,som06}).

Searches for the CGM have been conducted in ultraviolet (UV), but the 
association between galaxies and IGM absorbers still remains an unsettled
issue. Within impact distances of 
$r_\rho\lsim350$ kpc from the sight lines of background QSOs, 
all galaxies with $L>0.1L^*$ seem to 
have ``associated'' Ly$\alpha$ or other low ionized absorbers
(e.g., \citealt{chen08,wak09}). However, not all Ly$\alpha$ absorbers have an
associated galaxy found within 1$h^{-1}$ Mpc (e.g., \citealt{mor93,sto06}).
These absorbers could be either associated with faint galaxies that
are beyond the detection limits of current galaxy surveys or 
not related to any galaxies at all 
(some absorbers are detected in voids; \citealt{chen09, sto07}.) 
Also, it is possible that much of the injection 
into the
IGM occurred at high redshifts, and in the time that has subsequently
passed, substantial separations in space and velocity have developed
between the ejected matter and the galaxies from which it originated
\citep{tri06}.
The highly ionized
absorbers, \ovi\ absorbers in particular, may trace the CGM at 
high temperatures. Covering fraction of these absorbers 
varies with respect to galaxy luminosities and the impact distances
of these absorbers to their nearest galaxies
\citep{sto06, pro06, gan08, wak09}.

X-ray observations could provide complementary or even key constraints
on the existence and properties of the CGM. The bulk of the CGM, 
at least for galaxies with masses similar to
and higher than our Galaxy, is believed to be in a gaseous phase
at temperatures of $T\gsim 10^6$~K and mainly emits and absorbs
X-ray photons (e.g., \citealt{bir03}). However, because of
the low density and potentially low metallicity of the CGM, its X-ray emission,
except for that in galaxy clusters where the missing baryon problem
becomes less severe, is expected to be hard to detect (e.g., \citealt{han06}),
and the observational evidence is still absent (e.g., \citealt{ras09, and10}).

The high spectral resolution grating instruments aboard \chandra\ 
and \xmm\ now provide us with a new powerful tool --- X-ray
absorption line spectroscopy ---
to probe the low-surface brightness diffuse hot CGM. Unlike X-ray 
emission that is sensitive to the emission measure of the hot gas, 
absorption lines produced by helium- and hydrogen-like 
ions such as \ovii, \oviii, and \neix\  
directly probe their column densities,
which are proportional to the mass of the gas and are sensitive to its 
thermal, chemical, and kinematic properties.
Indeed, both the \chandra\ and \xmm\ grating observations of bright 
AGNs and X-ray binaries have been used to firmly detect and 
characterize the global hot gas in and around the Milky Way 
(e.g., \citealt{wang05, yao05, yao06, yao07, fang06, bre07}).
However, our location in the 
midst of the hot gas makes it hard to determine the presence of the CGM 
around the Galaxy; our previous investigations only yield tentative upper 
limits to the column densities of the CGM beyond the hot gas-rich 
Galactic disk (e.g., \citealt{yao08, yao09a}).

In this work, we search for absorption lines of \cvi, \nvii, \ovii, 
\oviii, and \neix\ produced in the CGM of intervening galaxies
along 12 AGN sight lines by extensively exploring 
\chandra\ grating observations. Because detection sensitivities of
current high resolution X-ray instruments are still limited,
to enhance the signal-to-noise (S/N) ratio, we stack
spectra with respect to group membership and luminosity properties 
of intervening galaxies. This stacking technique, 
as implemented in our previous paper \citep{yao09b}, enables
us to probe the absorbing gas to unprecedented low column densities.

The paper is organized as follows. In Section~\ref{sec:data},
we discuss the background AGNs and survey of intervening
galaxies, and we describe the data reduction process. 
We search for and measure the X-ray absorption lines produced in the CGM
in Section~\ref{sec:results}, and discuss the implications
of our results 
in Section~\ref{sec:dis}.

Throughout the paper, we adopt the Schechter luminosity function with
characteristic luminosity $L_B^*\simeq2\times10^{10}L_\odot$ 
(or absolute B magnitude $M_B^*\simeq-20.5$) 
and use the cosmology with $H_0=73~{\rm km~s^{-1}~Mpc^{-1}}$, 
$\Omega_{M}=0.3$, and $\Omega_{\Lambda}=0.7$.

\section{Background AGNs, Intervening galaxies, \chandra Observations,
	 and Data Reduction}
\label{sec:data}

We selected our background AGNs and their \chandra observations
using the following criteria.
For all AGNs observed with {\sl Chandra} grating instruments, 
we chose those sources
without intrinsic ($z\approx z_{\rm AGN}$)
soft X-ray emission/absorption lines reported in the 
literature. We also included those AGNs with intrinsic lines 
(e.g., \citealt{yaq03, mas03, roz04}; also see \citealt{mck07} and 
references therein), but showing no lines with 
$EW/\Delta EW$$\ge2\sigma$ ($EW$ is the 
equivalent width of a line) at wavelengths longer than 10 \AA\
in their \chandra\ spectra. 
We excluded observations with nonstandard configurations 
[e.g., observations that put target source outside of the S3 chip of the 
advanced CCD Imaging Spectrometer (ACIS) for calibration purposes] to avoid 
spectral resolution degradation.
We did not use several short ($<10$ ks) exposures of PKS~2155-304
that together contribute $\lsim10\%$ to the total spectral counts.
Among the candidate AGNs, we only used those sight lines along which there 
are $\geq$1 intervening galaxies at projected impact 
distances of $r_\rho\lsim$500 kpc. We used 500 kpc as a test limit, 
which is comparable to the size of regular galaxy groups 
(e.g., \citealt{tul87, hel00}). 
The \ovi\ and \ciii\ ultraviolet 
absorbers are found to have a median impact distance
of 350-500 kpc and a maximum distance of 800 kpc 
around $L^*$ galaxies \citep{sto06}.

\begin{figure*}
\plotone{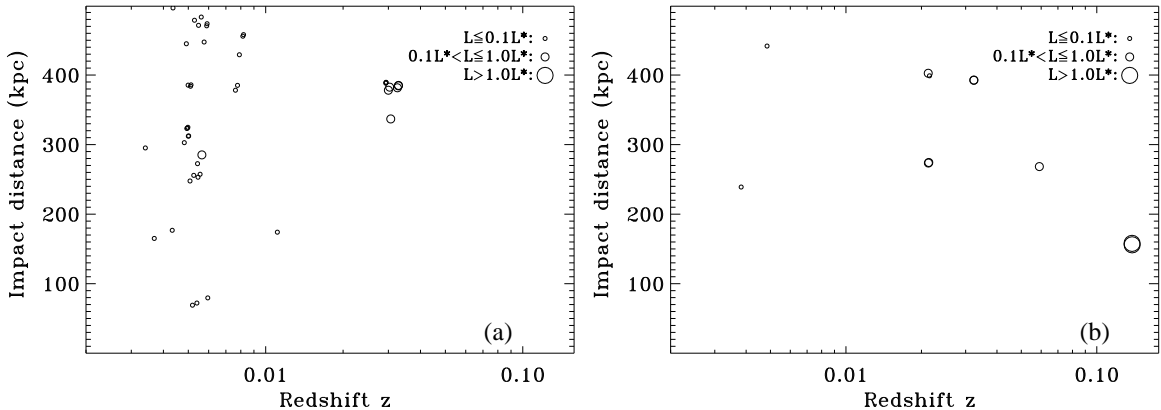}
\caption{Impact distances of intervening galaxies versus
	redshifts along the 3C~273 (a) and 
	the PG~1116+215 sight lines (b). }
\label{fig:interv}
\end{figure*}

The information on the intervening galaxies was extracted 
from the same galaxy database as used by \citet{sto06}.
This database was assembled from several galaxy redshift 
surveys and has been updated with SDSS DR7. Only those galaxies 
with reported B-band magnitudes were used, and 
the K-correction of \citet{fuk95} had been applied to obtain
the absolute magnitude of these galaxies.
Table~\ref{tab:obs} summarizes our selected AGNs, the number of 
{\sl Chandra} grating observations and total exposure time used in this
work, and the number of intervening galaxies along each sight line.
Table~\ref{tab:H1821+643}  
lists the intervening galaxies used in
this work. 
Figure~\ref{fig:interv} shows the intervening
galaxy distributions along the 3C~273 and PG~1116+215 sight 
lines for demonstration.

\begin{deluxetable}{lcrrr}
\tablewidth{0pt}
\tablecaption{\small Background AGNs, {\sl Chandra} observations,
and intervening galaxies \label{tab:obs}}
\tablehead{
Src. Name        & $z_{\rm AGN}$ & No. of Obs. & Exp. & No. of gal. \\
                 &	         &    & (ks) &    }
\startdata
H1821+643        & 0.297      & 5           &600 & 7(5)  \\
3C 273           & 0.158      & 17          &530 & 47(44) \\ 
PG 1116+215      & 0.176      & 1           &89  & 12(11) \\ 
PKS 2155-304     & 0.117      & 46          &1075& 14(13) \\
Ton S180         & 0.062      & 1           &80  & 3(3)  \\
PG 1211+143      & 0.081      & 3           &141 & 46(45) \\
Mrk 766 	 & 0.013      & 1           & 90 & 13(12)  \\
H1426+428	 & 0.129      & 3           & 184& 3(3)  \\
1H 0414+009 	 & 0.287      & 2           & 88 & 4(2)  \\
Mrk 509          & 0.034      & 1           & 59 & 1(1)  \\ 
IC 4329a 	 & 0.016      & 1           & 60 & 3(3)  \\ 
Fairall 9 	 & 0.047      & 1           & 80 & 1(1)  \\ 
\hline			                         
\multicolumn{2}{l}{Sub total:}&82           &3076& 154(143)   
\enddata
\tablecomments{
Values in parentheses indicate numbers of intervening 
galaxies with reported B-magnitudes.
Only these galaxies were used in this work.}
\end{deluxetable}

\begin{deluxetable*}{cccrccc}
\tablewidth{0pt}
\tabletypesize{\footnotesize}
\tablecaption{\small Intervening galaxies along the H1821+643 sight line
\label{tab:H1821+643}}
\tablehead{
RA (J2000) & Dec (J2000) & $z_g$ & $\Delta\theta$ & $b$ & $m_B$ & $L/L^*$ \\
(deg) & (deg) & & (arcsec) & (kpc) & (mag) & }
\startdata
 275.365 &  64.203 & 0.027519 &  9.02 & 300.9 & 20.60 & 0.006 \\
 275.010 &  64.315 & 0.027879 & 12.55 & 424.1 & 18.50 & 0.041 \\
 275.512 &  64.361 & 0.121537 &  1.19 & 176.0 & 18.30 & 1.502 \\
 275.403 &  64.357 & 0.170858 &  2.38 & 492.3 & 19.10 & 1.857 \\
 275.477 &  64.335 & 0.225603 &  0.60 & 164.8 & 19.50 & 2.920 \\
 275.520 &  64.348 & 0.284400 &  0.85 & 291.8 & 22.20 & 0.593 \\
\enddata
\tablecomments{$L/L^*$ is calculated in B band. Information of intervening galaxies along other sight lines are available online.}
\end{deluxetable*}

There are two detectors [the ACIS and the 
high resolution camera (HRC)] and two grating spectrographs 
[the high energy transmission grating (HETG) and
the low energy transmission grating (LETG)] 
aboard {\sl Chandra}~\footnote{Please refer to {\sl Chandra} Proposer's Observatory Guide
(http://asc.harvard.edu/proposer/POG/) for further information.}.
Observations with the LETG
collected the majority of the data used
in this work.

\begin{figure*}
\plotone{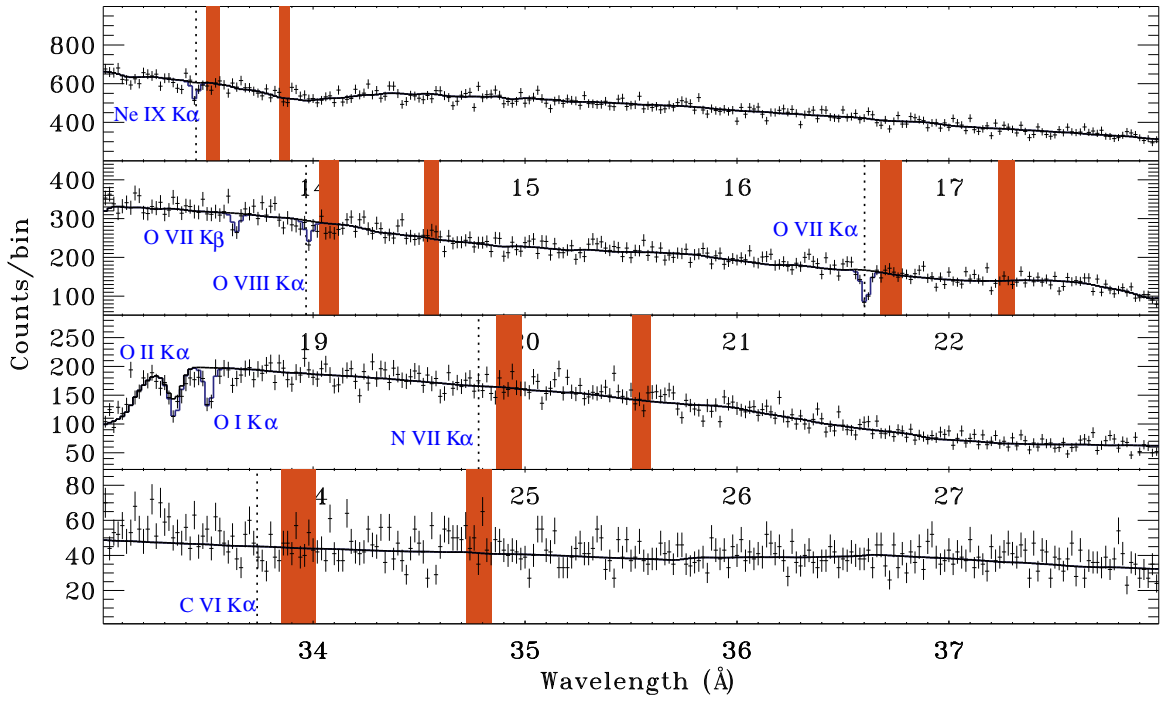}
\caption{Four ranges of the final spectrum of 3C~273. Blue lines and 
	labels mark locations of Galactic interstellar medium absorption
	(see Section~\ref{sec:absline} for further discussions).
	Red-shaded regions mark the spectral ranges of the expected	
	highly ionized 
	absorption lines produced in the CGM of intervening galaxies 
	in two groups (Figure~\ref{fig:interv}a); dotted 
	lines mark the corresponding rest-frame wavelengths of these 
	lines. The binsize is 20 m\AA\ in all spectra presented
	in this work. See text for details.}
\label{fig:3C273counts}
\end{figure*}
\begin{figure*}
\plotone{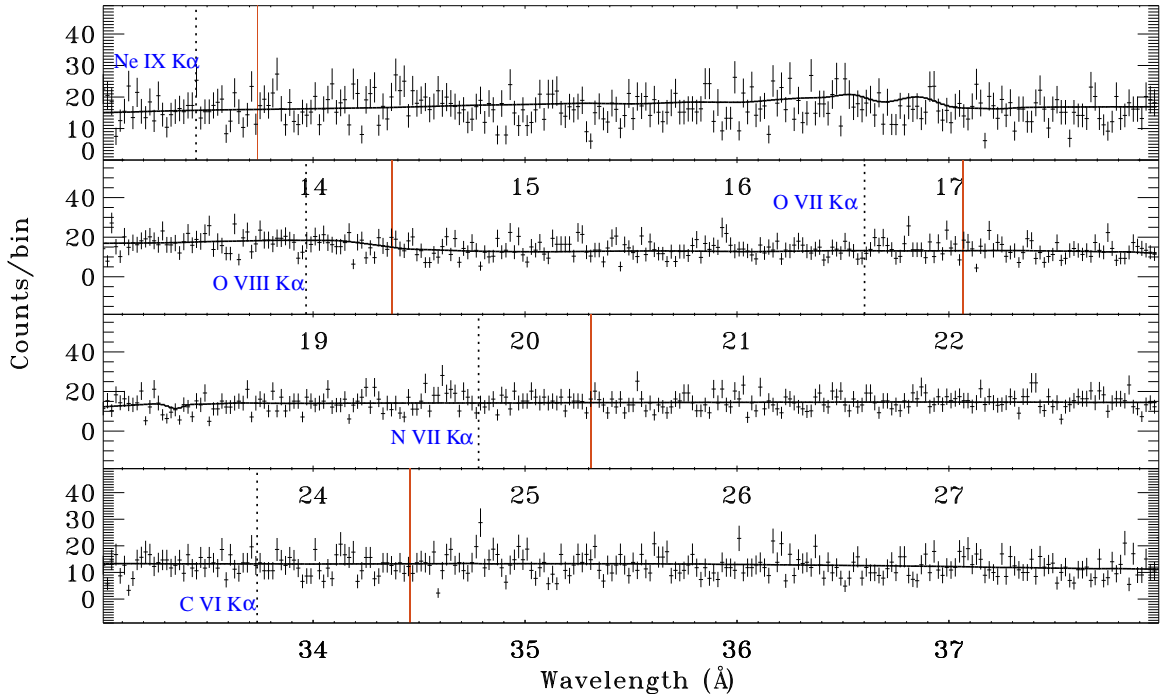}
\caption{Same as Figure~\ref{fig:3C273counts}, but for 
	PG~1116+215 sight line.}
\label{fig:PG1116counts}
\end{figure*}

We followed the same procedures as described in \citet{wang05}
and \citet{yao05, yao07} to perform the \chandra
data reduction.
Most observations of the first six sources listed in
Table~\ref{tab:obs} have been reported in \citet{yao09b}, and
the same spectra and the corresponding instrumental response 
files as obtained by \citet{yao09b} were used here. 
We extracted spectra and calculated 
the instrumental response files for those observations 
that have not been analyzed.
For each sight line with multiple observations, we also combined
spectra and corresponding response files to form a single 
coadded spectrum and response file. 
Figures~\ref{fig:3C273counts} and \ref{fig:PG1116counts} demonstrate the final 
spectra of 3C~273 and PG~1116+215,
and Table~\ref{tab:sn} summarizes S/N ratios of 
each AGN spectrum in 20-m\AA\ bins 
around the restframe wavelengths of the K$\alpha$ 
transitions of \neix, \oviii, \ovii, \nvii, and \cvi.

\begin{deluxetable}{lrrrrr}
\tablewidth{0pt}
\tablecaption{\small S/N ratios of AGN spectra
   around key lines.\label{tab:sn}}
\tablehead{
Src. Name & \neix & \oviii & \ovii & \nvii & \cvi}
\startdata
H1821+643        &  13.1 &   9.6 &   7.6 &   8.1 &   3.7 \\
3C 273           &  24.2 &  16.8 &  12.3 &  12.9 &   6.6 \\
PG 1116+215      &   4.1 &   4.1 &   3.8 &   3.7 &   3.5 \\
PKS 2155-304     &  45.3 &  37.9 &  30.4 &  31.0 &  21.6 \\
Ton S180         &   6.2 &   5.4 &   3.4 &   3.7 &   1.6 \\
PG 1211+143      &   5.3 &   4.4 &   3.3 &   3.5 &   1.5 \\
Mrk 766          &   5.3 &   3.3 &   2.4 &   2.7 &   $<$1 \\
H1426+428        &  10.8 &   6.3 &   4.4 &   4.1 &   $<$1 \\
1H 0414+009      &   3.4 &   2.0 &   $<$1 &   1.1 &   $<$1 \\
Mrk 509          &   5.4 &   3.1 &   2.1 &   2.0 &   $<$1 \\
IC 4329a         &   5.5 &   1.9 &   $<$1 &   1.3 &   $<$1 \\
Fairall 9        &   4.7 &   2.5 &   1.4 &   1.8 &   $<$1 
\enddata
\tablecomments{The restframe wavelengths of the K$\alpha$ transitions of 
   \neix, \oviii, \ovii, \nvii, and \cvi\ are 13.448 \AA, 18.967 \AA,
   21.602 \AA, 24.781 \AA, and 33.736 \AA, respectively \citep{ver96}.}
\end{deluxetable}

\section{Data analysis and results}
\label{sec:results}

We first search for the K$\alpha$ absorption lines of
\neix, \oviii, \ovii, \nvii, and \cvi\ produced in the CGM 
along individual sight lines. Most of the spectra of our
selected AGNs are of low counting statistics,
and only three of them, H1821+643, 3C~273, and 
PKS~2155-304, have sufficiently high S/N ratios to warrant such a search
(Figures~\ref{fig:3C273counts}-\ref{fig:PG1116counts} and 
Table~\ref{tab:sn}). 
Visual inspection reveals no convincing X-ray absorption line 
at the redshift of each intervening galaxy along any of these sight lines.

\begin{figure}
\plotone{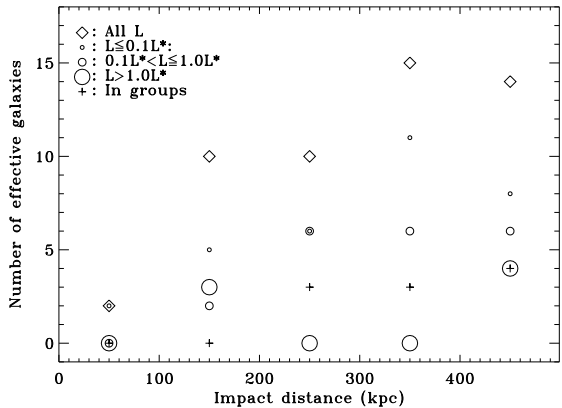}
\caption{The number of galaxies in five impact-distance bins for
	different galaxy categories. The numbers of all galaxies 
	(``All L'') in some distance bins are less than the sum of the
	numbers of individual 
	luminosities, which reflects the galaxy independent requirement in
	spectrum coadding. Please see text for details.}
\label{fig:gal_dis}
\end{figure}

\begin{figure*}
\plotone{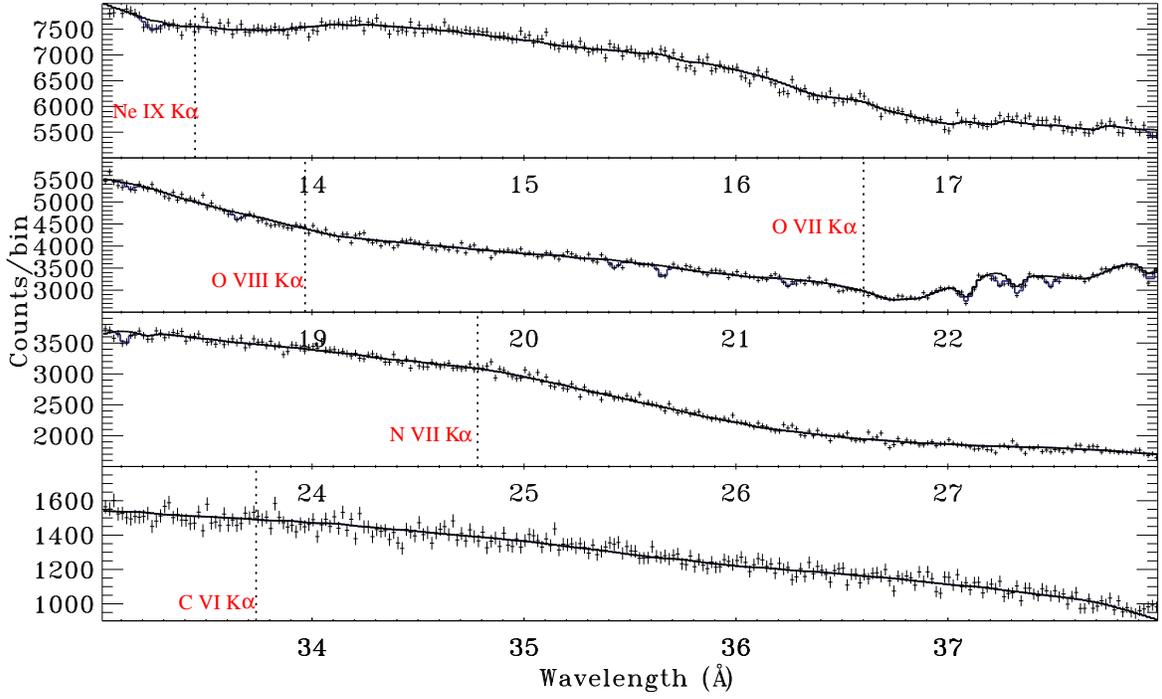}
\caption{Four ranges of the coadded spectrum sampling the
	intervening galaxies in groups along all sight lines.
	Model line profiles in blue superimposed on the data are 
	the Galactic absorptions contributed mainly 
	from the PKS~2155-304 sight line 
	(please refer to Figure~\ref{fig:3C273counts}); 
	dotted lines and 
	red labels mark positions of the expected absorption lines
	produced in the CGM of these galaxies.	
\label{fig:grp}}
\end{figure*}

\begin{figure*}
\plotone{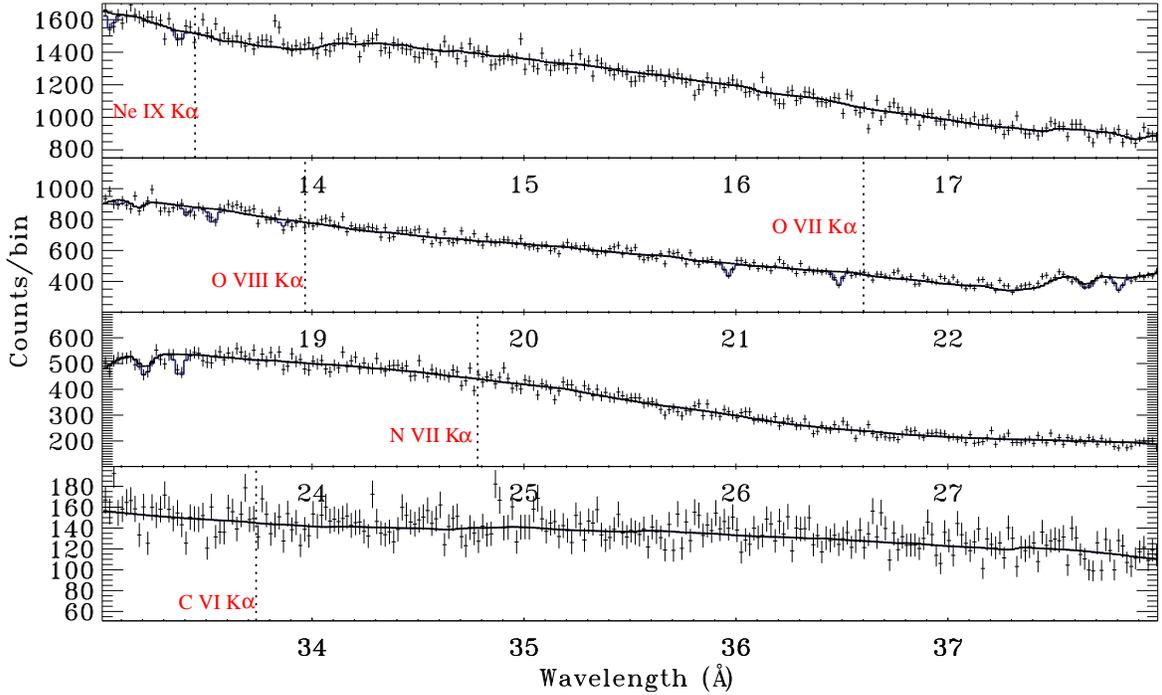}
\caption{Four ranges of the coadded spectrum sampling the
	intervening galaxies with luminosities
	of $0.1L^*<L\le L^*$ along all sight lines except for
	PKS~2155-304. 
	Model line profiles in blue superimposed on the data are
	the Galactic absorptions mainly contributed
	from the 3C~273 sight line 
	(please refer to Figure~\ref{fig:3C273counts}); dotted lines and 
	red labels mark positions of the expected absorption lines
	produced in the CGM of these galaxies. 
\label{fig:lum-1.0}}
\end{figure*}

We then search for the X-ray absorption lines produced in 
the CGM of the intervening galaxies from stacked spectra. 
We classify the galaxies 
into five categories, corresponding to galaxies in groups, galaxies with
luminosites of $L>L^*$, $0.1L^*<L\le L^*$, $L<0.1L^*$, and all galaxies.
We define a galaxy group along a sight line as consisting of three 
or more galaxies with line-of-sight velocity differences 
$c\Delta z_{ij}\le1000~{\rm km~s^{-1}}$, where
$c$ is the speed of light and $\Delta z_{ij}$ is the redshift 
interval of any two galaxies in the group.
Figure~\ref{fig:gal_dis} illustrates the number of galaxies in each
category as a function of impact distances.
We coadd the spectra with respect to redshifts of the 
intervening galaxies ($z_{\rm gal}$) in the same category.
Redshifts of the CGM absorption lines are expected to be the same 
as those of the intervening galaxies~\footnote{The 
redshift of the putative CGM of a galaxy 
could be offset from the $z_{\rm gal}$ by several hundred 
km~s$^{-1}$ (e.g., \citealt{sto06}), which, however, is unresolvable 
by current X-ray instruments. Please see Section~\ref{sec:val} for
further discussion.}. To properly conduct 
the coadding, along each sight line, we first blueshift the AGN spectrum 
and accordingly squeeze the instrumental response file by $z_{\rm gal}$,
the redshift of each intervening galaxy
\footnote{Please refer to \citet{yao09b} for 
procedures of blueshifting a spectrum and the
corresponding response file.}, and 
then stack them to form a single 
spectrum and response file for each of the five categories. 
To assure independence each shifted spectrum in coadding, for those 
galaxies along the same sight line and in the same category
but at very close redshifts that 
cannot be resolved by \chandra\ spectral resolution (e.g., 
$c\Delta z_{ij}\le1000~{\rm km~s^{-1}}$; we take the 
resolution of 1000~${\rm km~s^{-1}}$ as a conservative value), 
we use an average redshift $\overline{z_{\rm gal}}$ defined as
\begin{equation}
\overline{z{\rm_{gal}}} = 
\frac{\sum_{i=1}^n z^i_{\rm gal}/{r_\rho^i}^2}{\sum_{i=1}^n 1/{r_\rho^i}^2},
\end{equation}
where $z^i_{\rm gal}$ and $r_\rho^i$ are the redshift and the impact
distance of a galaxy, respectively. 
For instance, along the 3C~273 sight line, galaxies at only two redshifts
(Figure~\ref{fig:interv}a)
are used to form the coadded spectrum of the ``galaxies in groups'' 
category. Among the 143 intervening galaxies listed in 
Table~\ref{tab:obs}, only 29 are ``independent'' in the 
``all intervening galaxies'' category (Table~\ref{tab:sn:coadded}).
Since the PKS~2155-304 sight line dominates the spectral
counts of these coadded spectra (Table~\ref{tab:sn}), 
to avoid possible bias caused by a single sight line, 
we also obtain another five similar coadded spectra
but without contribution of the PKS~2155-304 sight line.
Figures~\ref{fig:grp}-\ref{fig:lum-1.0} 
show two examples of these coadded spectra.
Table~\ref{tab:sn:coadded} 
summarizes the S/N ratios of these coadded spectra
around the expected X-ray absorption lines and the 
number of the galaxies and the sight lines that these spectra are
sampling. 

\begin{deluxetable*}{lccccccc}
\tablewidth{0pt}
\tablecaption{\small S/N ratios of coadded spectra
   around the K$\alpha$ transitions of key ions 
   \label{tab:sn:coadded}}
\tablehead{
No. & Galaxies & Sight lines & \neix & \oviii & \ovii & \nvii & \cvi}
\startdata
1$^a$ & 10(7) & 6(5) & 86.7(36.9) &66.2(25.6) &54.5(19.2) &55.6(19.4) &38.6(10.7) \\
2$^b$ &  6(4) & 4(3) & 68.5(19.5) &49.9(12.5) &46.0(13.2) &39.5( 8.5) &29.6( 4.8) \\
3$^c$ & 15(12)& 8(7) & 88.0(38.9) &67.2(27.9) &54.8(21.0) &55.7(21.0) &38.9(12.1) \\
4$^d$ & 16(14)&11(10)& 76.5(41.2) &58.1(29.0) &46.4(21.4) &47.9(22.0) &32.5(12.0) \\
5$^e$ & 29(24)&12(11)& 113.(48.4) &85.1(34.1) &72.4(26.9) &69.2(25.3) &49.2(14.3) 
\enddata
\tablecomments{
	Values in parenthesis indicate without 
	contribution from the PKS~2155-304 sight line. \\
	$^a$ Sampling 10(7) intervening galaxies in groups 
	along 6(5) sight lines. \\
	$^b$ Sampling intervening galaxies with $L>L^*$. \\
	$^c$ Sampling intervening galaxies with $0.1L^*<L\le L^*$. \\
	$^d$ Sampling intervening galaxies with $L\le 0.1L^*$.\\
	$^e$ Sampling all intervening galaxies. }
\end{deluxetable*}

We measure the absorptions of the CGM from these coadded spectra.
Since all background AGN 
spectra have been shifted by the redshifts $z_{\rm gal}$ of the 
intervening  galaxies before
being coadded, the absorption lines produced in the CGM
are thus expected to be at the restframe 
wavelengths of these lines in the coadded spectra. 
Again, visual inspection reveals no convincing absorption line
in any of these coadded spectra, with or without contribution from the
PKS~2155-304 sight line (e.g., Figures~\ref{fig:grp}-\ref{fig:lum-1.0}).
Adding the Gaussian profiles at the restframe wavelengths of K$\alpha$ 
transitions of \neix, \oviii, \ovii, \nvii, and \cvi, we constrain 
the 95\% upper limits to the equivalent widths (EWs) of these absorption
lines. Replacing the Gaussian profiles with our absorption line model
{\sl absline}~\footnote{This model is similar to an analysis of 
the curve of growth.
Please refer to \citet{yao05} for a detailed description.} and
fixing the dispersion velocity to $b=50~{\rm km~s}^{-1}$ for the absorbing
medium, we also obtain the upper limits to the corresponding 
ionic column densities in the CGM, 
which are reported in Table~\ref{tab:results}.

\begin{deluxetable*}{l|cc|cc|cc|cc|cc}
\tablewidth{0pt}
\tablecaption{\small The 95\% Upper limits to EWs 
	of the expected key lines and the corresponding
	ionic column densities 
        \label{tab:results}}
\tablehead{
No. & \multicolumn{2}{c}{\neix} & 
\multicolumn{2}{c}{\oviii} & 
\multicolumn{2}{c}{\ovii} & 
\multicolumn{2}{c}{\nvii} & 
\multicolumn{2}{c}{\cvi}\\
& EW & log[N(cm$^{-2}$)] 
& EW & log[N(cm$^{-2}$)] 
& EW & log[N(cm$^{-2}$)] 
& EW & log[N(cm$^{-2}$)] 
& EW & log[N(cm$^{-2}$)] \\
& (m\AA) & & (m\AA)& &(m\AA)& &(m\AA)&& (m\AA)}
\startdata
1$^a$ & 0.99(2.71) & 14.99(15.57) & 0.77(2.55) & 14.53(15.01) & 0.41(2.60) & 14.25(14.96) & 0.99(4.78) & 14.34(15.04) & 0.73(6.57) & 14.06(14.96) \\
2$^b$ & 0.55(2.11) & 14.76(15.36) & 2.56(8.76) & 14.05(16.29) & 1.62(3.28) & 14.76(15.25) & 1.18(9.96) & 14.41(15.53) & 1.56(18.8) & 14.29(15.67) \\
3$^c$ & 0.87(2.19) & 14.96(15.35) & 0.70(3.42) & 14.41(15.18) & 0.39(2.50) & 14.20(14.86) & 1.18(4.11) & 14.52(14.96) & 0.67(7.03) & 13.93(15.02) \\
4$^d$ & 0.92(2.06) & 14.92(15.35) & 1.05(1.81) & 14.56(15.35) & 0.69(1.25) & 14.51(14.86) & 0.97(2.32) & 14.37(14.66) & 0.83(3.03) & 14.01(14.62) \\
5$^e$ & 0.64(2.17) & 14.71(15.40) & 0.96(2.26) & 14.61(14.98) & 0.41(1.08) & 14.16(14.51) & 0.74(3.19) & 14.20(14.96) & 0.49(5.27) & 13.80(14.87) 
\enddata
\tablecomments{$^{a-e}$ Measurements are obtained from the spectra,
	sorted to the same order as listed in 
	Table~\ref{tab:sn:coadded}.}
\end{deluxetable*}

\section{Discussion}
\label{sec:dis}

We have presented a search for X-ray absorption lines of the CGM along 
12 sight lines. The limited counting statistics of the X-ray spectra of the 
background AGNs usually doesn't warrant such a search. 
To improve the S/N ratios, we have stacked the AGN spectra with respect to 
different group and luminosity properties of the intervening galaxies. 
We do not detect any X-ray absorption line in the 
coadded spectra although the high
S/N ratios 
enable us to constrain the ionic column density to be as low as
$N_{\rm OVII}\lsim(1.6-6.3)\times10^{14}~{\rm cm^{-2}}$ 
(Table~\ref{tab:results}). In the following,
we first examine the reliability of our results (Section~\ref{sec:val})
and then discuss implications of these results for the galaxy feedback 
(Section~\ref{sec:feedback}) and for the origin of the absorption 
lines at $z\simeq0$ routinely detected in spectra of many AGNs 
(Section~\ref{sec:absline}).

\subsection{Validity of our data analysis procedures}
\label{sec:val}

Here we address two crucial questions regarding to the analysis
procedures employed in this work: 
(1) Can our stacking restore the line significance
if there were X-ray absorptions produced in the CGM at 
various redshifts? (2) Could the absorption signal be 
severely blurred if the CGM absorptions are of several hundred 
km~s$^{-1}$ velocity offset
from those of galaxies? 
To answer these questions, we run Monte-Carlo simulations to 
obtain detection rates of the LETG for three different scenarios.

\begin{figure}
\plotone{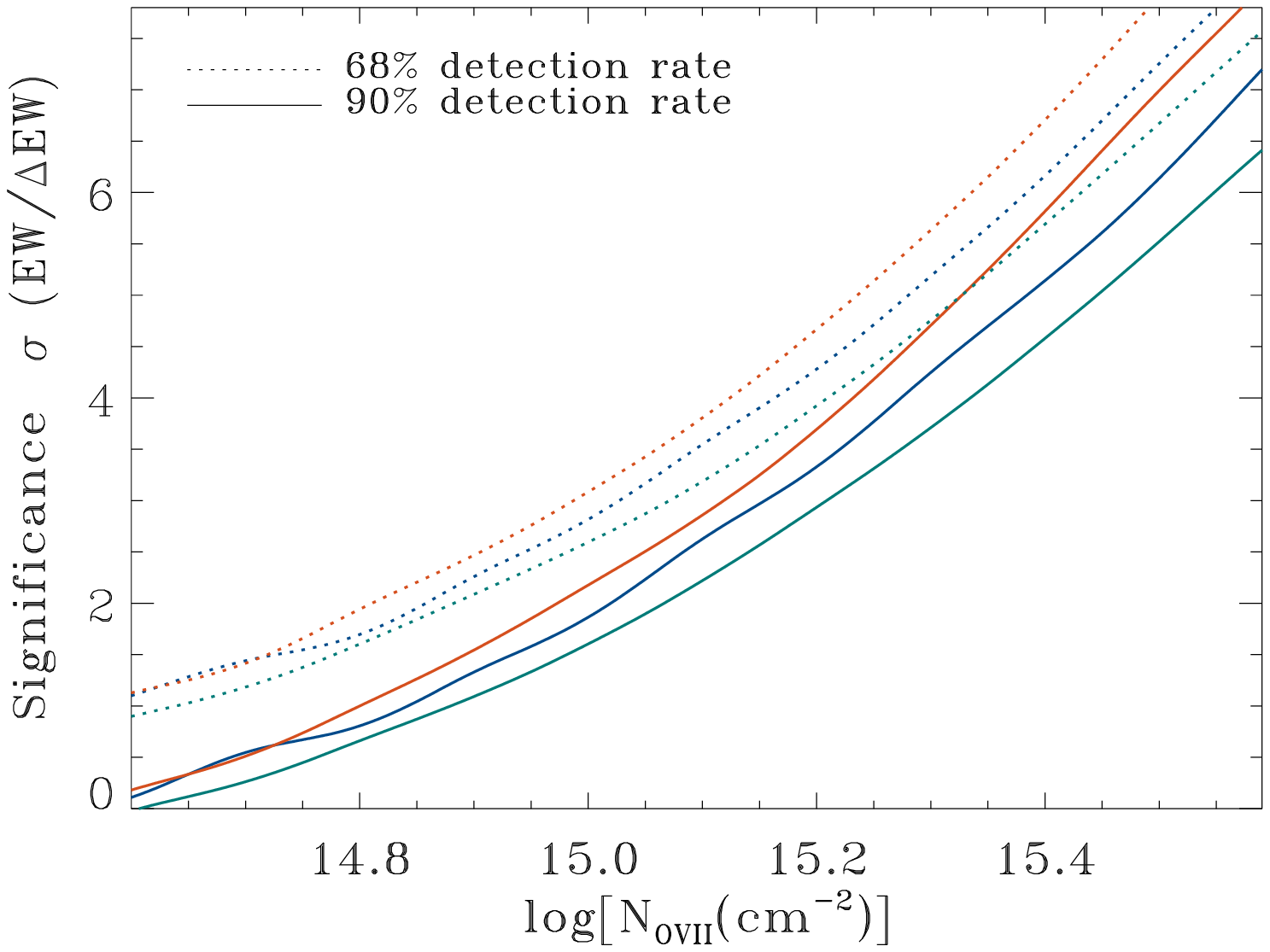}
\caption{The 68\% and 90\% possibilities of detecting the 
	\ovii\ K$\alpha$ 
	absorption line at certain significance levels or higher for
	various \ovii\ column densities.
	Blue curves indicate the results from individual spectra,
	each containing 2000 counts per 20-m\AA\ bin around the line.
	Red curves indicate results obtained from shift-coadded spectra, 
	that each contains ten spectra with 200 counts 
	per bin around the line. Ten absorption lines in the ten spectra
	are at redshifts evenly distributed from 0.01 to 0.1 to
	mimic various redshifts of the intervening galaxies.
	Green curves indicate the similar results
	to those of red curves, except for that the lines 
	in each set of the ten spectra are of random 
	velocity offsets from those intervening galaxies. 
	Please see text for details.
}
\label{fig:simulation}
\end{figure}

First, we simulate the probabilities of detecting an 
absorption line from a single spectrum.
Since the LETG collects most of the photons for the spectra used in
this work (Section~\ref{sec:data}) and the S/N ratios of the stacked
spectra are 40-50 around the \ovii\ K$\alpha$ absorption line
(Table~\ref{tab:sn:coadded}), 
we use the response file of the LETG to simulate spectra with 
spectral counts of 2000 per 20-m\AA\ bin around 21.6 \AA. We then measure the 
significance ($EW/\Delta EW$) of a simulated line for different 
input column densities ($N_{\rm OVII}$) and a fixed Doppler
dispersion velocity
of $v_{\rm b}=50~{\rm km~s^{-1}}$. We repeat these simulations 10,000 times 
and obtain the 68\% and 90\% detection rates, which describe
the probabilities of detecting the \ovii\ line at certain significance 
levels or higher for various $N_{\rm OVII}$ 
(Figure~\ref{fig:simulation}). 
For instance, the simulations indicate that for an assumed 
$N_{\rm OVII}=10^{15}~{\rm cm^{-2}}$, 
the 68\% and 90\% of the detections are at $\gsim$3.1 $\sigma$ 
and $\gsim$2.4 $\sigma$.

Second, we check the validity of our shifting and coadding procedures.
In this scenario, we first simulate a set of ten spectra, each with spectral 
counts of 200 per bin around the \ovii\ K$\alpha$ line. In these ten
spectra, the lines, all characterized with the same $N_{\rm OVII}$ and
$v_{\rm b} =50~{\rm km~s^{-1}}$,
are placed at ten redshifts distributed evenly between 0.01
and 0.1, to mimic various redshifts of the intervening galaxies.
We then use the same procedures as used in the actual data analysis to
blueshift these spectra by the corresponding redshifts, 
coadd them to form a single spectrum, and then measure
the line significance. As in the first
scenario, we repeat the whole process 10,000 times to obtain the
68\% and 90\% detection rates for different $N_{\rm OVII}$ inputs, 
which are also presented in 
Figure~\ref{fig:simulation}. It is interesting to note that, 
for a given $N_{\rm OVII}$ and with the same detection possibility, 
a line in the shift-coadded spectrum is expected to be detected
at a higher significance level than in a single spectrum with a similar
S/N ratio. This slightly enhanced significance is caused by
the higher spectral resolution ($\lambda/\Delta\lambda$)
of the LETG at longer wavelengths and channel rebinning when spectra are
blueshifted. This comparison 
validates our stacking procedures used in this work.
 
Finally, we examine the effect of the possible 
velocity offsets between the CGM absorptions and the intervening galaxies.
We run similar simulations as in the second scenario, except that
in each set of ten spectra, the absorption lines have random 
velocity offsets from the redshifts of the intervening galaxies. The 
offset velocities follow a normal distribution with a standard deviation of 
$200~{\rm km~s^{-1}}$ from each redshift \citep{sto06}.
Simulations indicate that in the high column density end 
(e.g., $\log [N_{\rm OVII} ({\rm cm^{-2}})]\gsim15.4$),
the detection significance, though could be discounted, is still high. 
In the low column density end (e.g., $\log [N_{\rm OVII} ({\rm
    cm^{-2}})]\lsim15.2$), to which our stacked spectra is approaching, 
the detection significance is not of much differences 
(Figure~\ref{fig:simulation}). In short, velocity offsets of several hundred
km~s$^{-1}$ between intervening  galaxies and the surrounding CGM would not
significantly blur the putative absorption signals 
in our stacked spectra.

\subsection{Implications for the missing baryon state}
\label{sec:feedback}

As discussed in Section~\ref{sec:intro}, galaxy feedback could be
in various forms for galaxies on different mass/luminosity scales and at
different evolutionary stages. The CGM is expected to 
contain important information about the galaxy feedback and could also be
a significant baryonic reservoir of individual dark matter halos. 
Taking $N{\rm _{OVII} \lsim10^{15}~cm^{-2}}$ and assuming that
the CGM is uniformly distributed around galaxies, we can estimate the
total mass contained in the CGM as 
$M_{\rm CGM}\lsim1\times(\frac{0.5}{f_{\rm
      OVII}})\times(\frac{0.3A_\odot}{A})\times(\frac{R}{500~
  {\rm kpc}})^2\times10^{11}M_\odot$,  
where $f_{\rm OVII}$, $A$, and $R$ are the ionization fraction of 
\ovii, metallicity, and the radial distance of the CGM, respectively.
In contrast, the hot CGM is expected to contain 
$\gsim3\times10^{11}~M_\odot$ baryons for Milky Way type galaxies and 
a typical galaxy group \citep{mcg08}.
These results thus indicate that the bulk of the CGM is unlikely to
reside in a chemically enriched warm-hot phase 
(at temperatures ranging from $10^{5.5}-10^{6.3}$ K) 
where our X-ray absorption line spectroscopy is sensitive. 

Our results, together with existing measurements of various species 
in other wavelength bands, can be used to constrain the state and/or 
extent of the CGM, hence its effect on galaxy 
evolution. One possibility is that the bulk of the baryon matter that 
was initially associated with the dark matter has been expelled out to 
large regions possibly on scales greater than individual group halos. 
Such stratification of baryon matter relative to dark matter halos 
may be a natural result of galactic feedback, initially via superwinds 
from starbursts and possibly AGNs (e.g., \citealt{mo04}) 
and later maintained by relatively 
gentle, but long-lasting energy injection from continuing star formation 
and by evolved stars (e.g., Type Ia SNe; \citealt{tang09a}). The exact 
physical/chemical state and extent of the stratified region is still 
uncertain, depending on the efficiency of the energy outputs from the 
starbursts and the star formation history as well as the environment 
of the galaxies.

While in this work the intervening galaxies are located at projected 
impact distances of 100-500 kpc along the selected sight lines
(e.g., Figure~\ref{fig:interv}), the ``associated'' Ly$\alpha$ and \ovi\ 
absorbers have also been detected at similar distances around galaxies 
(e.g., \citealt{sto06, wak09}). The gas traced by \ovi\ absorbers 
($T \sim 10^{5.3}$-$10^{5.7}$ K, if collisionally ionized) can account 
for up to 10\% of the baryon mass associated with individual dark matter
halos in the present Universe 
\citep{tri00, dan08, wak09}.
But recent work indicates that a large fraction of the \ovi\ absorbers
could be due to photoionization of cool gas clouds (e.g., \citealt{tri08}).
Therefore, for a remaining component to account for the missing baryons, 
its temperatures must be too high (e.g., $\gsim10^{6.3}$ K) 
and/or its metal abudance is too low 
($<0.1A_\odot$) to avoid the detection in X-ray absorption lines. 

While the metal 
abundance of the CGM is hardly constrained, the commonly accepted value
is about 0.2 $A_\odot$, as indicated by a few measurements of cool 
high-velocity clouds ($\approx 10^4$ K) around the Galaxy 
(e.g., \citealt{lu94, tri03, shu09}) and by the 
modeling of quasar absorption lines (e.g., \citealt{kee05}).
However, some studies have found circumgalactic material with
substantially higher metallicities
(e.g., \citealt{jen05, ara06}).
One may expect that the hot component would also have
higher metal abundances, as indicated by measurements for the medium 
in numerous galaxy clusters and a few well-observed groups
(e.g., \citealt{de01,pra07}). 

A high temperature and low density CGM component is 
predicted in numerical simulations, particularly for the evolution 
of Milky Way type galaxies and group environments \citep{tang09a, kim09}.
The high temperature is also consistent with 
observations of the large-scale hot intragroup medium, 
which seems to be concentrated around galaxies and is always at
temperatures higher than $\sim10^{6.8}$ K \citep{sun09}. A significant
portion of the CGM may be in the hot phase hiding from our detection; the
gas is too hot to produce detectable neon, oxygen, nitrogen, or carbon
absorption lines.
The observed \ovi\ absorptions can arise from either
photo-ionization of embedded cool clouds or collisional ionization at their
interfaces with the hot components (e.g., \citealt{wak09}). 
How well the hot CGM is mixed with the cool clouds and the relative
fractions of the CGM in the two phases likely depend on both the masses of
individual galaxies and the richness of the environment. 

Although our present investigation cannot constrain 
the CGM on scales of $\lsim$ 100 kpc around galaxies, our recent studies 
of the Galactic halo/CGM suggest that the observed \ovii\ absorption 
arises primarily in regions around the Galactic disk/bulge 
on scales of $\lsim10$ kpc \citep{yao08, yao09a}. 
Additional constraints have been discussed in a more recent work 
by \citet{and10}, who conclude that the hot gas cannot explain the missing
baryon matter on galactic halo scales. This conclusion complements the
constraints obtained in this paper primarily on larger scales. 

Properties of the CGM could vary from one galaxy to another. This
variation may have been revealed by the different column densities and the
non-unit covering factor of the \ovi\ absorbers 
(e.g., \citealt{sto06, wak09}). In this work, we group galaxies with
respect to different luminosities. The upper limits to the 
column densities of the X-ray absorbers and to the total mass of the
CGM thus should be regarded as average values for these galaxies.

In short, we find little evidence for the warm-hot gas in the temperature 
range of a few times $10^5$-$10^6$ K to account for a significant fraction of
the missing baryons in the vicinity of galaxies. 
The bulk of the CGM likely exists in cool clouds (e.g., \citealt{wak09}) or 
in a hot gas ($T\gsim10^{6.5}$ K).

\subsection{Implication for absorption lines at $z\simeq0$}
\label{sec:absline}

The results presented here also have strong implication for the
origin of absorption lines at $z\simeq0$ as demonstrated in 
Figure~\ref{fig:3C273counts}. Highly ionized absorption 
lines at $z\simeq0$, in particular the \ovii\
K$\alpha$ at 21.602 \AA, have been observed in all 
AGN spectra as long as the spectra have sufficiently high S/N ratios
(e.g., \citealt{fang06, bre07}).
The plausible locations of these absorbers include the hot interstellar 
medium (ISM) around the Galactic disk
(e.g., \citealt{yao05, yao07, yao08, yao09a}), the large-scale Galactic 
halo (e.g., \citealt{ras03, fang06, bre07}),
and intra-group gas in the Local Group
(e.g., \citealt{nic02, wil05}). There are no clear boundaries 
between the Galactic disk and the Galactic halo and between 
the Galactic halo and the intra-group gas, but the scale sizes of them
are expected to be several kpc, $\sim$10-100 kpc, and $\sim$100-1000 kpc,
respectively. This work probes the CGM located 
between the Galactic halo and intra-group gas.
Since the selected intervening galaxies
have impact distances of 100-500 kpc, our sight lines do not sample
much of galactic disks and halos. If the absorption lines at $z\simeq0$
are mainly produced in the intra-group gas, similar lines
are also expected for intervening galaxy groups
(e.g., Figures~\ref{fig:interv}a). However, we only obtain 
tightly constrained upper limits 
to the ionic column densities, in particular 
$N_{\rm OVII}\le10^{15}~{\rm cm^{-2}}$ 
(the first row of Table~\ref{tab:results}),
which, compared to the $z\sim0$ absorption 
$N_{\rm OVII}\simeq10^{16}~{\rm cm^{-2}}$ (e.g., \citealt{wil05, wil07}), 
indicate that the contribution of the intra-group gas, if 
it exists at all, must be
$\lsim10\%$. 
This conclusion is consistent with the constraint obtained
by comparing the X-ray
absorption with emission measurements \citep{fang06, yao07}
and with the non-detection of the expected spatial preferences of
the gas along the Local Group geometry \citep{bre07}, 
as well as with the differential absorption line spectroscopy based 
on different depths of sight lines \citep{yao08, yao09a}.

\acknowledgements

This work was partly supported by NASA
grant NNX08AC14G, provided to the University of Colorado to support
data analysis and scientific discoveries related to the Cosmic Origins
Spectrograph on the Hubble Space Telescope. YY, QDW, and TMT  
appreciate financial backing for this work provided by NASA ADP grants 
NNX10AE86G, NNX10AE85G, and NNX08AJ44G respectively.



\begin{thebibliography}{}
\bibitem[Anderson \& Bregman (2010)]{and10} Anderson, M. E., \& Bregman,
  J. N. 2010, \apj, in press
\bibitem[Aracil \etal (2006)]{ara06} Aracil, B., \etal 2006, \mnras, 367, 139
\bibitem[Birnboim \& Dekel (2003)]{bir03}Birnboim, Y., \& Dekel, A. 2003, MNRAS, 345, 349 
\bibitem[Bregman \& Lloyd-Davies (2007)]{bre07}Bregman, J., \& Lloyd-Davies, E. 2007, \apj, 669, 990 
\bibitem[Bregman \etal (2009)]{bre09} Bregman, J., \etal 2009, arXiv:0906.4993
\bibitem[Cattaneo \etal (2009)]{cat09} Cattaneo, A, \etal 2009, Nature, 460, 213
\bibitem[Cen \& Ostriker (1999)]{cen99}Cen, R., \& Ostriker, J. 1999, \apj,514, 1
\bibitem[Chen \& Tinker (2008)]{chen08} Chen, H.-W., Tinker, J. L. 2008,
\apj, 687, 745
\bibitem[Chen \& Mulchaey (2009)]{chen09} Chen, H.-W., Mulchaey, J. S. 2009, \apj, 701, 1219
\bibitem[Dav\'e \etal (1999)]{dav99}Dav\'e R., \etal 1999, \apj, 511, 521
\bibitem[Dav\'e \& Oppenheimer (2007)]{dav07}Dav\'e R., \& Oppenheimer, B. D. 2007, MNRAS, 374, 427 
\bibitem[Danforth \& Shull (2008)]{dan08}Danforth, C. W., \& Shull, J. M. 2008, \apj, 679, 194
\bibitem[De Grandi \& Molendi (2001)]{de01} De Grandi, S., \& Molendi, S. 2001, \apj, 551, 153
\bibitem[Dehnen \& Binney (1998)]{deh98}Dehnen, W., \& Binney, J., 1998, MNRAS, 294, 429 
\bibitem[Everett \etal (2008)]{eve08} Everett, J. E., \etal 2008, \apj,
  674, 258
\bibitem[Fang \etal (2006)]{fang06} Fang, T., \etal 2006, \apj, 644, 174 
\bibitem[Fukugita \etal (1995)]{fuk95} Fukugita, M., \etal 1995, \pasp, 107, 945
\bibitem[Gilmore (2008)]{gil08}Gilmore, G. 2008, Science 322, 1476 
\bibitem[Ganguly \etal (2008)]{gan08} Ganguly, R., Cen, R., Fang, T., \& Sembach, K. 2008, \apj, 678, L89
\bibitem[Hansen \& Sommer-Larsen (2006)]{han06}Hansen, S., \& Sommer-Larsen, J. 2006, \apj, 653, L17 
\bibitem[Helsdon \& Ponman (2000)]{hel00} Helsdon, S. F., \& Ponman,
  T. J. 20000, \mnras, 315, 356
\bibitem[Hoekstra \etal (2005)]{hoe05} Hoekstra, H, \etal 2005, \apj, 635, 73
\bibitem[Jenkins \etal (2005)]{jen05} Jenkins, E. B., Bowen, D. V., Tripp,
  T. M., \& Sembach, K. R. 2005, ApJ, 623, 767
\bibitem[Keeney \etal (2005)]{kee05} Keeney, B. A., \etal 2005, \apj, 622, 267
\bibitem[Kim \etal (2009)]{kim09} Kim, J., Wise, J, H., \& Abel, T. 2009, \apjl, 694L, 123, 
\bibitem[Komatsu \etal (2009)]{kom09} Komatsu, E., \etal 2009, \apjs, 180, 330
\bibitem[Lu \etal (1994)]{lu94} Lu, L., Savage, B. D., \& Sembach, K. R. 1994, \apjl, 437, L119
\bibitem[Mac Low \& Ferrara (1999)]{mac99}Mac Low , M.-M., \& Ferrara, A. 1999, ApJ, 513, 142 
\bibitem[Mason \etal (2003)]{mas03} Mason, K. O., \etal 2003, \apj, 582, 95
\bibitem[Morris \etal (1993)]{mor93} Morris, S. L., \etal 1993, \apj, 419, 524
\bibitem[McGaugh (2008)]{mcg08}McGaugh, S. S. 2008, IAUS, 244, 136
\bibitem[McKernan \etal (2007)]{mck07}McKernan, B.,Yaqoob, T., \& Reynolds, C. S., 2007, \mnras, 379, 1359
\bibitem[Mo \& Mao \etal (2004)]{mo04} Mo, H. J. \& Mao S. 2004, \mnras, 353, 829
\bibitem[Navarro \etal (1996)]{nav96}Navarro J. F., \etal 1996, ApJ, 462, 563 
\bibitem[Nicastro \etal (2002)]{nic02} Nicastro, F., \etal 2002, ApJ, 573, 157
\bibitem[Pratt \etal (2007)]{pra07} Pratt, G. W., \etal 2007, A\&A, 461, 71
\bibitem[Primack (2009)]{pri09}Primack, J. P. 2009, NJPh, 11, 5029
\bibitem[Prochaska \etal (2006)]{pro06} Prochaska, J. X., Weiner, B. J., Chen, H.-W., \& Mulchaey, J. S. 2006, \apj, 643, 680
\bibitem[Rasmussen \etal (2003)]{ras03} Rasmussen, A., Kahn, S. M., \& Paerels, F. 2003, in The IGM/Galaxy Connection, ed. J. L. Rosenberg \& M. E. Putman (ASSL Vol. 28: Dordrecht: Kluwer), 109	
\bibitem[Rasmussen \etal (2009)]{ras09} Rasmussen, J., \etal 2009, \apj, 697, 79
\bibitem[R\'o\.za\'nska \etal (2004)]{roz04} R\'o\.za\'nska, A., \etal 2004, \apj, 600, 96
\bibitem[Sheth \etal (2001)]{she01}Sheth R. K., \etal 2001, MNRAS, 323, 1 
\bibitem[Shull \etal (2009)]{shu09}Shull, J. M., \etal 2009, \apj, 699, 754 
\bibitem[Sommer-Larsen (2006)]{som06}Sommer-Larsen, J. 2006, \apj, 644, 1 
\bibitem[Springel \etal (2005)]{spr05} Springel, V., \etal 2005, Nature,
  435, 629
\bibitem[Stocke \etal (2006)]{sto06} Stocke, J. T., \etal 2006, \apj, 641, 217 
\bibitem[Stocke \etal (2007)]{sto07} Stocke, J. T., \etal 2007, \apj, 671, 146
\bibitem[Strickland \etal (2002)]{str02}Strickland, D. K., \etal 2002, \apj, 568, 689 
\bibitem[Sun \etal (2009)]{sun09} Sun, M., et al. 2009, ApJ, 693, 1142
\bibitem[Swartz \etal (2006)]{swa06} Swartz, D. A., \etal 2006, \apj, 647, 1030 
\bibitem[Tang \etal (2009a)]{tang09a}Tang, S., \etal 2009a, \mnras, 392, 77 
\bibitem[Tang \etal (2009b)]{tang09b} Tang, S., \etal 2009b, \mnras, 398, 1468
\bibitem[Tripp \& Savage (2000)]{tri00}Tripp, T. M., \& Savage, B. D. 2000, \apj, 542, 42 
\bibitem[Tripp \etal (2006)]{tri06}Tripp, T. M., \etal 2006, \apj, 643, L77 
\bibitem[Tripp \etal (2003)]{tri03}Tripp, T. M., \etal 2003, \aj, 125, 3122
\bibitem[Tripp \etal (2008)]{tri08}Tripp, T. M., \etal 2008, ApJS, 177, 39
\bibitem[Tully (1987)]{tul87} Tully, R. B. 1987, \apj, 321, 280
\bibitem[Verner \etal (1996)]{ver96} Verner, D. A., \etal 1996, ADNDT, 64, 1 
\bibitem[Wakker \& Savage (2009)]{wak09}Wakker, B., \& Savage, B. 2009, \apjs, 182, 378
\bibitem[Wang \etal (2005)]{wang05} Wang, Q. D., \etal 2005, \apj, 635, 386
\bibitem[Williams \etal (2005)]{wil05} Williams, R., \etal 2005, ApJ, 631, 856
\bibitem[Williams \etal (2007)]{wil07} Williams, R., \etal 2007, ApJ, 665, 247
\bibitem[Yao \& Wang (2005)]{yao05}Yao, Y., \& Wang, Q. D., 2005, \apj, 624, 751 
\bibitem[Yao \& Wang (2006)]{yao06}Yao, Y., \& Wang, Q. D., 2006, \apj, 641, 930 
\bibitem[Yao \& Wang (2007)]{yao07}Yao, Y., \& Wang, Q. D., 2007, \apj, 658, 1088 
\bibitem[Yao \etal (2008)]{yao08}Yao, Y., \etal 2008, \apj, 672, L21 
\bibitem[Yao \etal (2009a)]{yao09a}Yao, Y., \etal 2009a, \apj, 696, 1418 
\bibitem[Yao \etal (2009b)]{yao09b}Yao, Y., \etal 2009b, \apj, 697, 1784 
\bibitem[Yaqoob \etal (2003)]{yaq03}Yaqoob, T., \etal 2003, \apj, 582, 105

\end{thebibliography}
\end{document}